\documentclass[useAMS,usenatbib]{mn2e}

\usepackage{graphicx}
\usepackage{color}

\newcommand{\ud}{\,\mathrm{d}}

\title[Tendrils and voids in GAMA]
	{Galaxy and Mass Assembly (GAMA): Fine filaments of galaxies detected within voids}

\author[M. Alpaslan et al.]
	     {Mehmet Alpaslan$^{1,2}$, Aaron S.G.~Robotham$^2$, Danail Obreschkow$^2$, Samantha Penny$^3$, \newauthor
	      Simon Driver$^{1,2}$, Peder Norberg$^4$ Sarah Brough$^7$, Michael Brown$^3$, Michelle Cluver$^5$, \newauthor
	      Benne Holwerda$^6$, Andrew M. Hopkins$^7$, Eelco van Kampen$^8$, Lee S. Kelvin$^9$, \newauthor
	      Maritza A. Lara-Lopez$^7$, Jochen Liske, $^8$, Jon Loveday$^{10}$, Smriti Mahajan$^{11}$, \newauthor
	      Kevin Pimbblet$^{3,12,13}$\\
$^1$SUPA, School of Physics and Astronomy, University of St Andrews, North Haugh, St Andrews, Fife, KY16 9SS, UK\\
$^2$International Centre for Radio Astronomy Research, 7 Fairway, The University of Western Australia, Crawley, Perth,\\
 Western Australia 6009, Australia\\
$^3$School of Physics, Monash University, Clayton, Victoria 3800, Australia\\
$^4$Institute for Computational Cosmology, Department of Physics, Durham University, South Road, Durham, DH1 3LE, UK\\
$^5$Department of Astronomy, University of Cape Town, Private Bag X3, Rondebosch, 7701, South Africa\\
$^6$European Space Research and Technology Centre, Keplerlaan 1, 2200 AG Noordwijk, The Netherlands\\
$^7$Australian Astronomical Observatory, PO Box 915, North Ryde, NSW 1670, Australia\\
$^8$ESO, Karl-Schwarzschild-Str. 2, D-85748 Garching bei M\"{u}nchen, Germany\\
$^9$Institut f\"{u}r Astro- und Teilchenphysik, Universit\"{a}t Innsbruck, Technikerstra{\ss}e 25, 6020 Innsbruck, Austria\\
$^10$Astronomy Centre, University of Sussex, Falmer, Brighton, BN1 9QH, UK\\
$^{11}$School of Mathematics and Physics, The University of Queensland, St Lucia 4072, Australia\\
$^{12}$Department of Physics, University of Oxford, Denys Wilkinson Building, Keble Road, Oxford OX1 3RH, UK\\
$^{13}$Department of Physics and Mathematics, University of Hull, Cottingham Road, Hull, HU6 7RX, UK\\}

\pagerange{\pageref{firstpage}--\pageref{lastpage}}\pubyear{2014}

\begin{document}

\label{firstpage}

\maketitle

\begin{abstract}
Based on data from the Galaxy and Mass Assembly (GAMA) survey, we report on the discovery of structures that we refer to as `tendrils' of galaxies: coherent, thin chains of galaxies that are rooted in filaments and terminate in neighbouring filaments or voids. On average, tendrils contain 6 galaxies and span 10 $h^{-1}$ Mpc. We use the so-called line correlation function to prove that tendrils represent real structures rather than accidental alignments. We show that voids found in the SDSS-DR7 survey that overlap with GAMA regions contain a large number of galaxies, primarily belonging to tendrils. This implies that void sizes are strongly dependent on the number density and sensitivity limits of a survey. We caution that galaxies in low density regions, that may be defined as `void galaxies,' will have local galaxy number densities that depend on such observational limits and are likely higher than can be directly measured.
\end{abstract}

\section{Introduction}

The $ > 1 h^{-1}$ Mpc environment of a galaxy in a void is sparse relative to a galaxy in a dense cluster, making such systems ideal for studying the evolution of galaxies independent of environmental processes. Being extremely underdense regions, voids are an easily recognised feature of the Cosmic Web. They span between 20 to 50 $h^{-1}$ Mpc and have a typical galaxy density that is 10\% less than that surrounding a typical field galaxy \citep{Pan2012}. The few galaxies that do exist in voids are subject to dynamics that are unique to these underdense regions \citep{Blumenthal1992,Sheth2004}. Voids serve as tools for constraining cosmological parameters, or for testing the accuracy of large cosmological simulations, as shown in \citet{Dekel1994,Lavaux2010,Park2012}. The shapes of voids are defined by the baryonic matter that surrounds them, namely filaments and walls. These features form the easily recognised large scale structure (LSS) in the Universe we see today. Galaxy surveys have conclusively shown, through a variety of filament finding methods (e.g., \citealp{Doroshkevich2004,Pimbblet2005,Colberg2007,Forero-Romero2009,Aragon-Calvo2010,Murphy2011,Smith2012,Alpaslan2013}), that galaxies tend to coalesce in large linear structures that can span up to $60$ to $100 h^{-1}$ Mpc or beyond \citep{Gott2005}. There is little consensus, however, as to an exact quantitative description of the properties of LSS, be it filaments or voids. 

Early studies of voids \citep{Joeveer1978,Gregory1978,Kirshner1981,deLapparent1986} have been supplemented by more recent large galaxy redshift surveys \citep{Colless2001,Abazajian2009} that provide comprehensive and complete pictures of voids, complemented by numerous void-finding techniques (e.g., \citet{El-Ad1997,Hoyle2002,Aragon-Calvo2010a}. The Void Galaxy Survey (VGS, \citealp{Kreckel2012}) is one such effort that has identified galaxies in voids from the SDSS-DR7 data. \citet{Beygu2013} have recently identified the system VGS\_31, which is comprised of three galaxies in a void that appear to be aligned linearly. The system is linked together by a single massive cloud of H{\sc I} gas. \citet{Rieder2013} have recently shown that such structures can be accurately reproduced in simulations, however; the existence of substructures in voids has been previously predicted (including `low-density ridges' of galaxies within voids \citealp{VandeWeygaert1993,Gottloeber2003}). The detection of such prominent structure in a void is an indication that they merit further study. Observationally, this means designing and conducting surveys that are expansive, deep, and highly spectroscopically complete on all scales.

GAMA \citep{Driver2009,Driver2011} is an ongoing spectroscopic survey that spans three equatorial fields centred on the equator at $\alpha$ = 9h (G09), $\alpha$ = 12h, (G12) and $\alpha$ = 14.5h (G15) and two southern fields centred on $\alpha$ = 34 deg, $\delta$ = -7 deg (G02) and $\alpha$ = 345 deg and $\delta$ = -32.5 deg (G23). Each field measures $12 \times 5$ deg$^2$, out to $m_r < 19.8$ mag and $z \leq 0.5$. The survey is highly spectroscopically complete ($ > 98\%$) and probes the galaxy stellar mass function down to $M_* \approx 10^8 M_{\odot}$, giving us an unprecedented view into the realm of low mass galaxies in the nearby Universe. This, combined with up to 1050 targets per square degree, makes GAMA an ideal body of data for searching for LSS spanning many orders of magnitude in galaxy stellar mass.

\citet{Alpaslan2013} have used the GAMA survey to look for LSS in the nearby ($z \leq 0.213$) Universe. The GAMA Large Scale Structure Catalogue (Alpaslan et al. 2013; hereafter referred to as the GLSSC) classifies a subset of 45,542 GAMA galaxies as belonging to filaments, voids, and introduces a third population of galaxies lying in what we refer to as tendrils. These are described in detail in this paper. Tendrils are coherent structures typically containing up to 5 or 6 galaxies that span roughly 10 $h^{-1}$ Mpc and are rooted in filaments, either connecting to other filaments or terminating in voids. In Section 2 we briefly discuss the creation of the GLSSC and the data on which it is based. Section 3 describes tendrils in detail, focusing on their structure and how they are located with respect to voids. Throughout this paper, we use $H_0 = 100$ kms$^{-1}$ Mpc$^{-1}$, $\Omega_M = 0.25$, and $\Omega_{\Lambda} = 0.75$, and $h = H0/100 \mathrm{km}^{-1} \mathrm{s}^{-1}\mathrm{Mpc}^{-1} = 1$ is assumed when calculating absolute magnitudes.

\begin{figure*}
	\centering
	\includegraphics[width=0.9\textwidth]{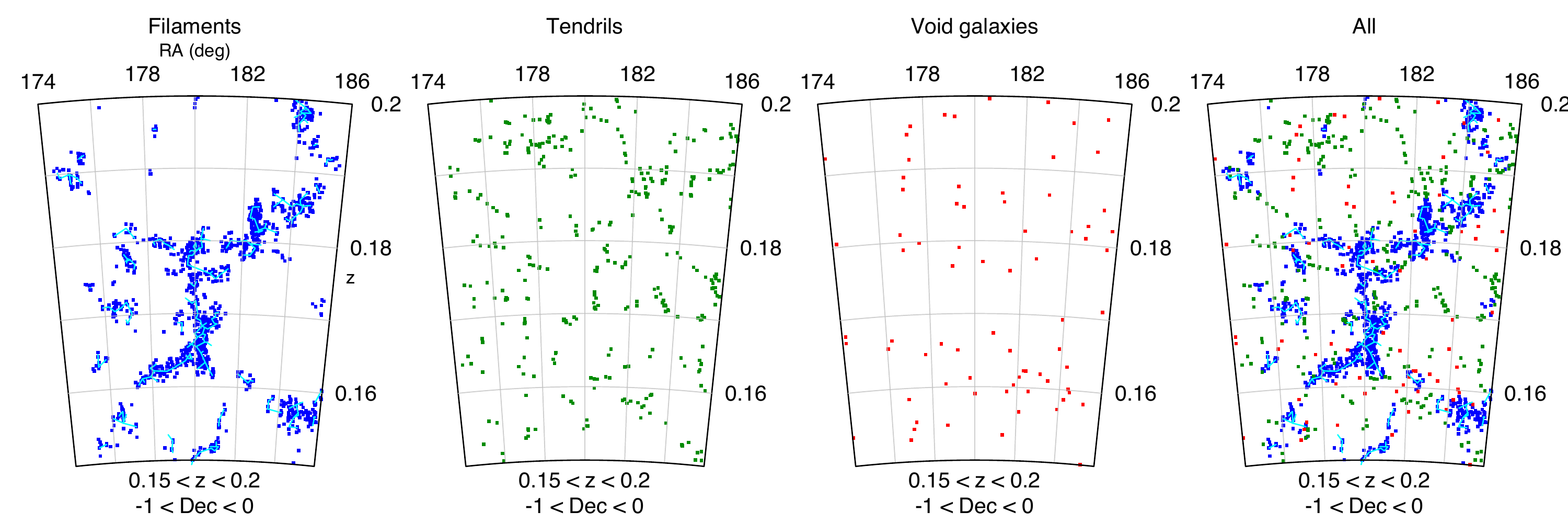}
	\caption{A section of the G12 field with different galaxy populations shown in each panel. From left to right the populations shown are galaxies in filaments with the filament minimal spanning tree (blue and cyan respectively); galaxies in tendrils (green); galaxies in voids (red); and all three populations in their respective colours.}
	\label{fig:strip}
\end{figure*}

\section{LSS in GAMA}

The GLSSC classifies 45,542 galaxies with $0 \leq z \leq 0.213$ and $M_r \leq -19.77$ mag in the G09, G12, and G15 fields as belonging to different kinds of LSS using a modified minimal spanning tree algorithm. The GLSSC establishes three hierarchies in LSS: filaments and voids, and a third intermediate population referred to as tendrils. Filaments are large structures of galaxies and groups that exist in highly overdense regions, with intricate geometries (not just straight linear structures). The largest filaments are often better described as large `complexes' or conglomerations of filaments that intersect at a massive node. Void galaxies are found in underdense regions that surround filaments and show very little clustering signal at all scales beyond 5 $h^{-1}$ Mpc. Tendrils are substructures of galaxies that exhibit coherent linear structure, but do not have the same levels of overdensity that filaments do; they appear to be rooted in filaments, branching into other filaments or terminating in voids. In Alpaslan et al. (2013) we find that for a sub-sample of galaxies with $M_* \geq 10^{10.61} M_{\odot}$ and $m_r < 19.4$ mag, tendril galaxies contain up to one quarter of the total stellar mass of galaxies in the GLSSC.

Below, we provide an abridged description of the method used to generate the GLSSC but for full details, we refer the reader to Alpaslan et al. (2013). The algorithm uses a three pass approach, where the first step is to identify primary filaments composed of galaxy groups \citep{Robotham2011}. This is done by generating a minimal spanning tree (MST) across all groups in a volume-limited sample with galaxy coordinates converted into a comoving Euclidean space. Any links greater than a distance $b$ are trimmed. This value is chosen such that at least 90\% of groups with $L_* \geq 10^{11} L_{\odot}$ are found in filaments. For the second pass, all galaxies at a distance $r$ from the nearest filament are associated with that filament, and referred to as `filament galaxies' alongside galaxies within filament groups. This population represents the most prominent and easily identifiable LSSs. The final pass involves generating an MST on all galaxies \emph{not} in filaments, and trimming any links greater than a length $q$. Galaxies left in the remaining structures are identifed as tendril galaxies, and all isolated galaxies are referred to as void galaxies. The parameters $r$ and $q$ are selected such that the two point correlation function $\xi(R)$ of void galaxies shows no signal at large distances (given by $R$); quantitatively this is done by selecting $q$ such that $\int R^2 \xi(R) \ud R$ is minimised. Tendril galaxies are therefore defined such that they must be at a distance $r$ away from the nearest filament galaxy, and at no further than $q$ from the nearest tendril galaxy. The GLSSC is generated with $b = 5.75 h^{-1}$ Mpc; $r = 4.12 h^{-1}$ Mpc; and $q = 4.56 h^{-1}$ Mpc. This defition of a void galaxy is different to the usual method of defining void galaxies based on their presence witin a low density region, which allows for possible clustering. We also generate a set of LSS catalogues for a set of GAMA mock catalogues, which are described in \citet{Robotham2011} and \citep{Merson2013}. These mock catalogues are 9 mock light cones, generated by populating haloes from the Millennium Simulation \citep{Springel2005} with the \textsc{Galform} semi-analytic model \citep{Bower2006} and are designed to match the geometry and $r$-band survey luminosity function of the GAMA survey \citep{Loveday2012}. In Alpaslan et al. (2013) we show that there is excellent agreement between observed filamentary structures in the GLSSC and filaments extracted from the mock catalogues.

\section{Tendrils}

In Figure \ref{fig:strip}, we show the different galaxy populations in a $12 \times 11$ deg$^2$ slice of the G12 field, with $0.15 \leq z \leq 0.2$. The left panel displays all filament galaxies in blue, with the MST links shown on top in cyan. Tendril galaxies are shown in the next panel, void galaxies in the third, and all four populations of galaxies are shown together in the rightmost panel. This final panel shows that tendrils generally branch off from filaments and penetrate into voids, often bridging between two distinct filaments. Most importantly, tendrils arc across ranges of right ascension and declination as well as redshift, suggesting that they are not statistical alignments by chance or caused by redshift space distortions, but real structures - a claim that will be substantiated in Section 3.1.

\subsection{Filamentarity}

We can quantify the filamentarity, defined as the linearity of structure, of the filament, tendril and void galaxies using their line correlation $l(r)$. This correlation is an estimator of spatial statistics introduced by \citet{Obreschkow2013} to characterize the cosmic density field $\delta(\vec{r})$ or its Fourier transform (FT) $\hat{\delta}(\vec{k})$. Since the isotropic power-spectrum $p(k)\propto k^{-2}\!\int d\phi\,d\theta|\hat{\delta}(\vec{k})|^2$ exhaustively exploits the information contained in the amplitudes $|\hat{\delta}(\vec{k})|$ of a homogeneous and isotropic field, \citeauthor{Obreschkow2013} chose to investigate the residual information in the phase factors $\hat\epsilon(\vec{k})\equiv\hat{\delta}(\vec{k})/|\hat{\delta}(\vec{k})|$. The inverse FT $\epsilon(\vec{r})$ has a vanishing two-point correlation $\xi_2(r)$; thus the lowest non-trivial correlations of $\epsilon(\vec{r})$ are three-point correlations. The line-correlation $\l(r)$ is defined as a suitably normalised version of the isotropic three-point correlation of $\epsilon(r)$ for three points on a straight line, separated by $r$. Geometrically, it turns out that $l(r)$ measures the degree of straight filamentarity on length scales $r$ in about the same way that $\xi_2(r)$ measures the clustering on scales $r$. Figure 3 in \citet{Obreschkow2013} displays $l(r)$ for a series of density fields and illustrates how $l(r)$ can distinguish linear over-densities from spherical ones, unlike $\xi_2(r)$ and more robustly than the traditional three-point correlation.

Given the non-cartesian volume of the GAMA cones and the varying number of galaxies in each sample, we chose to measure the filamentarity via the excess line correlation $\Delta l(r)=[l(r)-l_0(r)]\sqrt{f}$. Here, $l(r)$ is the line correlation of the galaxies, taken as points of equal mass, truncated to distances larger than $400 h^{-1}$ Mpc and fitted in a cubic box of $215~h^{-1}$ Mpc side length. The reference function $l_0(r)$ is the line correlation of an equal number of points, randomly distributed in an equivalent survey volume. The factor $f$ is the volume fraction of the GAMA cones within the cubic box. The term $\sqrt{f}$ ensures that $\Delta l(r)$ is approximately independent of the volume of the cubic box (according to Section 3.4 of \citep{Obreschkow2013}).

\begin{figure}
	\centering
	\includegraphics[width=0.4\textwidth]{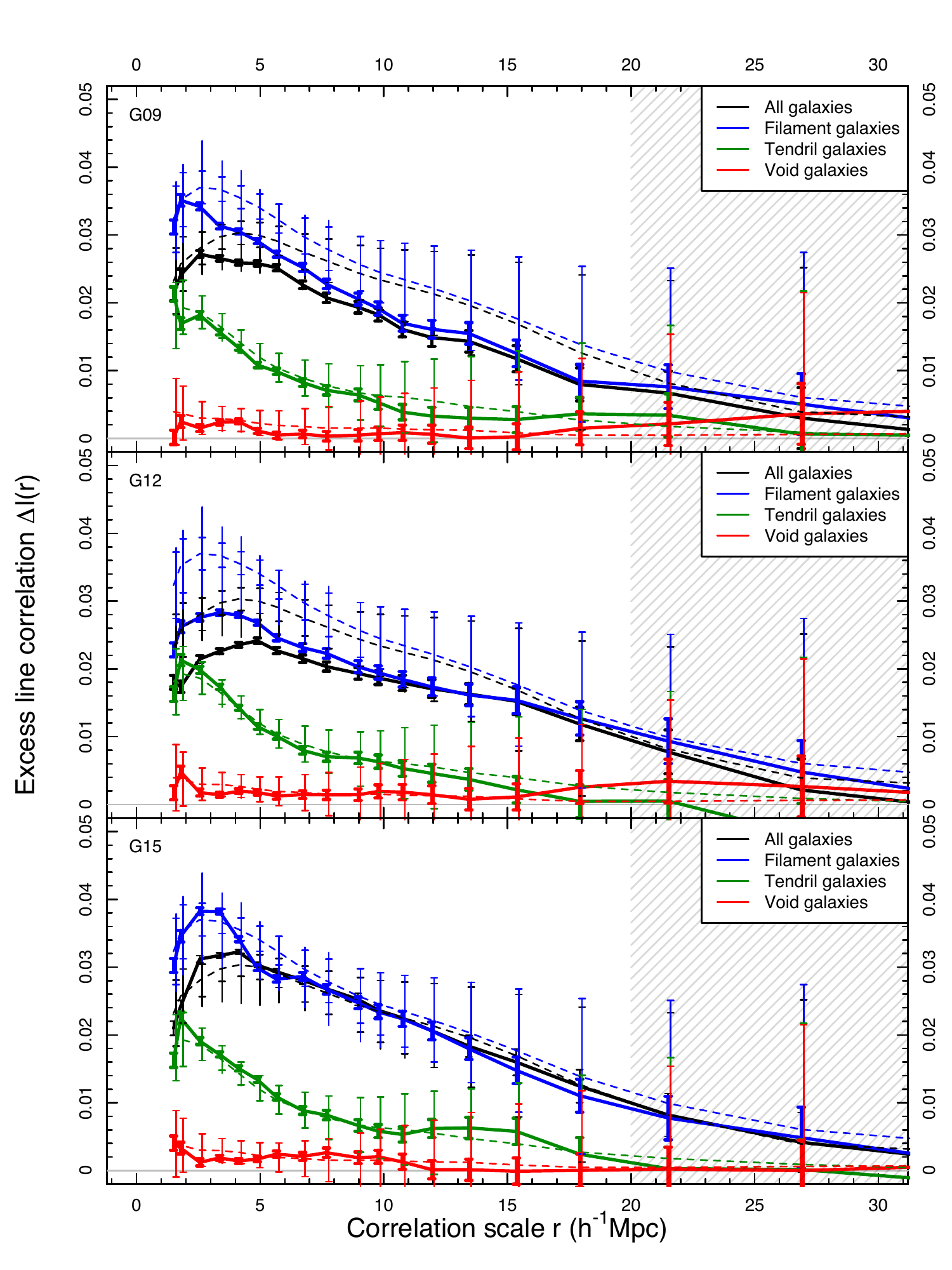}
	\caption{Excess line correlation $\Delta l(r)$ shown for filament (blue), tendril (green) and void galaxies (red), as well as all galaxies combined (black) across all three equatorial GAMA fields. The dashed lines show the median $\Delta l(r)$ for the GAMA mock catalogues, for which we show both the formal $\Delta l(r)$ error as typical error bars, and the spread between mock lightcones as solid lines with no arrowheads. The function $\Delta l(r)$ roughly measures the probability of finding three equidistant points separated by $r$ on a straight line, in a density field that is analogous to GAMA but stripped of all two-point correlation. Shaded regions show lengths where $\Delta l(r)$ is unreliable due to the limited declination range of the GAMA cones ($\sim40~h^{-1}$ Mpc on average).}
	\label{fig:linecorr}
\end{figure}

We calculate $\Delta l(r)$ separately for filament, tendril and void galaxies, as well as all galaxies combined. This calculation is performed individually for each of the three equatorial GAMA fields  as well as for the GAMA mock catalogues. Figure \ref{fig:linecorr} shows the functions $\Delta l(r)$ of the observed data (solid lines) with measurement uncertainties (1$\sigma$ error bars). These uncertainties arise from the limited number of independent Fourier modes in the finite survey volume. They do \textit{not} include cosmic variance due to correlated LSS. In turn, dashed lines and their error bars represent the median and standard deviation of $\Delta l(r)$ of the nine GAMA mock catalogs. These standard deviations naturally include both, uncertainties in the measurements of $\Delta l(r)$ and cosmic variance in the survey volume. Since these error bars are much larger than those of the observed data, cosmic variance is the dominant uncertainty in $\Delta l(r)$ across all considered scales.

Observed and simulated filaments, tendrils and voids show very similar values for $\Delta l(r)$ on all scales across the three GAMA fields. In all cases, filaments alone exhibit a higher correlation $\Delta l(r)$ than all galaxies together. Void galaxies show very little excess line correlation at all scales, indicating that for the magnitude limited sample in the GLSSC, void galaxies are free from structure. Tendrils show a clear intermediate line correlation, leaving no doubt that those structures contain real filamentarity well beyond that of a random point set. This example effectively demonstrates the usefulness of higher-order statistical estimators such as $l(r)$ to characterize LSS.

\subsection{Relationship with voids}

In the GLSSC, a void galaxy is defined as a galaxy that is at least $4.56\;h^{-1}$ Mpc away from the nearest galaxy that belongs to a tendril (and tendril galaxies themselves must be at least $ 4.12\;h^{-1}$ Mpc away from the nearest filament), so they exist in the most underdense regions of the GAMA fields. We do not determine where voids are, or their sizes, only the galaxies that exist in very underdense regions. 

Recently, \citet{Pan2012} have released a catalogue of void galaxies obtained from the SDSS-DR7 data using a voidfinder similar to the one introduced by \citet{El-Ad1997}. In \citet{Pan2012}, voids are identified within a volume limited subsample of 120606 SDSS galaxies with $M_r < -20.09$ mag and $z < 0.107$. Galaxies are first determined to be within the field or not, using the third nearest neighbour distance $d_3$ and the standard deviation of this distance $\sigma_{d_3}$. Any galaxy with $d = d_3 + 1.5 \sigma_{d_3} > 6.3 h^{-1}$ Mpc is considered to belong to a cosmic filament or a cluster and is referred to as a `wall' galaxy. All other galaxies are classified as field galaxies and are removed from the sample. Wall galaxies are gridded into cells of 5 $h^{-1}$ Mpc, and all empty cells are considered to be possible centres of voids. A sphere is grown from each cell until it is bounded by four wall galaxies, and any two spheres with more than 10\% overlap are considered to belong to the same void. A sphere must have a radius of at least 10 $h^{-1}$ Mpc to be considered a void.

\begin{figure}
	\centering
	\includegraphics[width=0.4\textwidth]{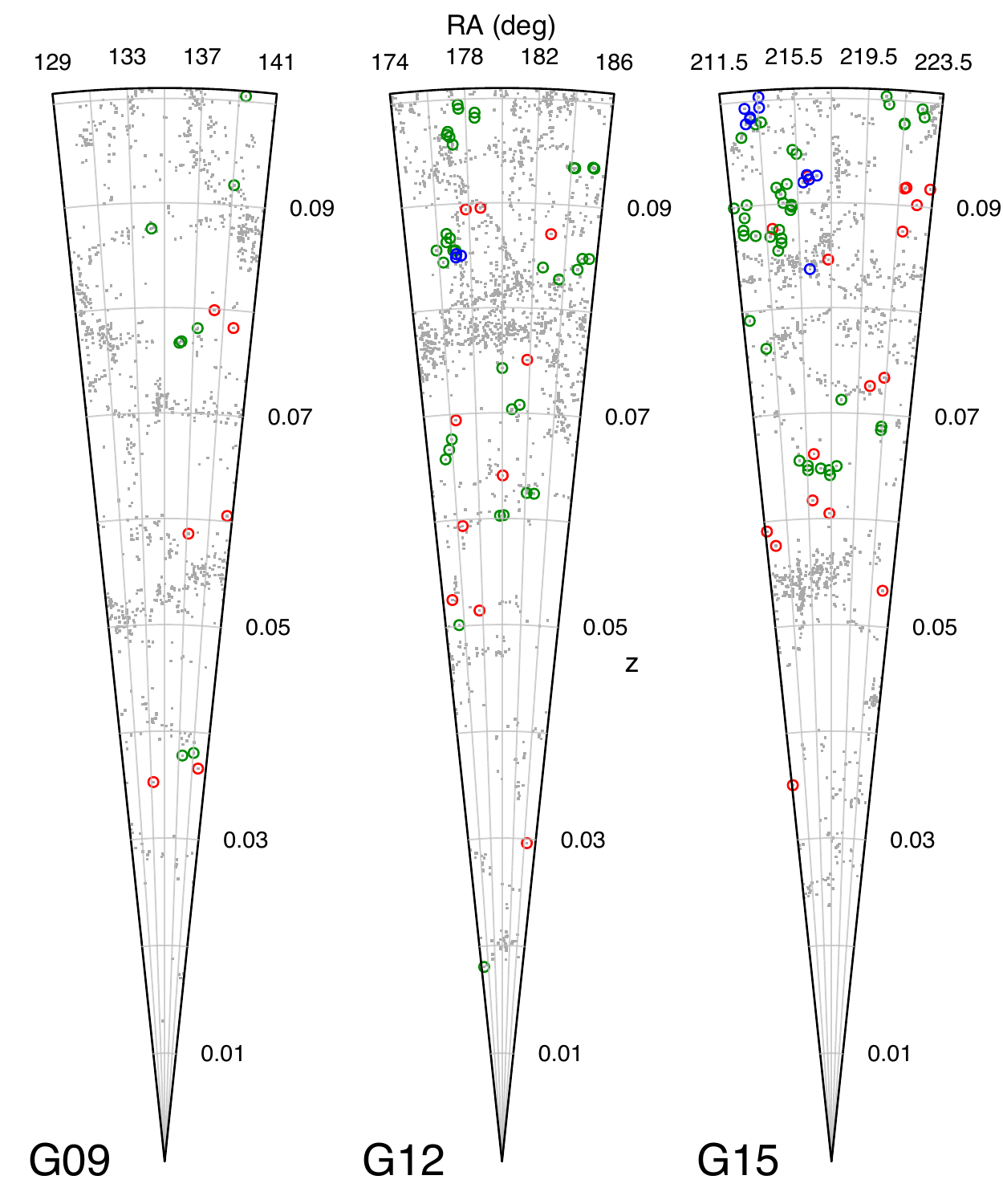}
	\caption{Each circled galaxy in this figure is a galaxy in the GLSSC that is located inside a void identified in SDSS-DR7 data: blue circles represent galaxies in filaments, green circles represent galaxies in tendrils, and red circles represent galaxies in voids. We note that we show the full $5^{\circ}$ declination for each field in this figure, resulting in increasing projection effects at higher redshifts.}
	\label{fig:voids_filaments}
\end{figure}

Using the SDSS-DR7 void catalogue from \citet{Pan2012}, we identify 1130 galaxies with $0.001 \leq z \leq 0.1$ and $m_r < 19.8$ within voids that lie in the three equatorial GAMA fields, excepting the $-1^{\circ} < \delta < 0^{\circ}$ region of G09. We define the voids that are used to compare to GAMA data using the maximal sphere radius parameter of each void from the \citet{Pan2012} catalogue, and find 132 galaxies in the GLSSC that lie within the inner two thirds of these voids. These galaxies are displayed in Figure \ref{fig:voids_filaments}, where we show all galaxies in the GLSSC for the three equatorial fields, and circle the 132 galaxies galaxies that are matched to the 1130 GAMA galaxies in the SDSS voids. The circles are coloured according to the structures those galaxies are determined to be in the GLSSC: blue circles are galaxies in filaments, green circles are galaxies in tendrils, and red circles are galaxies in voids. We find that only 25\% of GLSSC galaxies with $M_r < -19.77$ mag found in SDSS voids are galaxies that we also identify to be isolated, with the vast majority (64\%) aligned along fine tendrils, and a further 11\% associated with filaments. For a subsample of 79 galaxies with $M_r < -20.06$ mag, 15\% are filament galaxies, 61\% in tendrils and 24\% in voids.  Penny et al., in prep, will examine, in detail, the properties of the 1130 GAMA galaxies in SDSS voids.

% \begin{table}
% 	\centering
% 	\begin{tabular}{lccc}
	
% 	 &Filaments&Tendrils&Voids\\
% 	 \hline
% 	 $M_r < -19.77$&11\%&64\%&25\%\\
% 	 $M_r < -20.06$&15\%&61\%&24\%\\
% 	 \hline
	 
% 	 \end{tabular}
	 
% 	 \caption{Percentage of galaxies, grouped by large scale environment, matched with galaxies in voids located in GAMA fields, for two different absolute magnitude cuts.}
% 	 \label{table:percentages}
	 
% \end{table}

\section{Discussion and summary}

Using the GLSSC, we have identified a new class of structures in addition to groups, filaments and voids. Tendrils exhibit a coherent linear structure, but do not have the same levels of overdensity that filaments do, and either connect filaments or are rooted in one filament and terminate in a void. In the GLSSC we identify 14150 galaxies that lie within such structures (compared to only 2048 void galaxies and 29503 galaxies in filaments). Tendril galaxies contain 23.8\% of all stellar mass for a magnitude and mass limited sample of the GLSSC; however they are tenuous enough to be difficult to detect using structure finders that rely on density smoothing, or only consider galaxies in groups. These methods will not fully take into account less massive galaxies outside of the most high density regions.

As structures, tendrils appear to be morphologically distinct from filaments in that they are more isolated and span shorter distances. On average, a tendril will contain just under 6 galaxies and measure roughly 10 $h^{-1}$ Mpc in length. When examining their filamentarity using the line correlation function $l(r)$ we find that they are distinguished from filaments by having an intermediate excess line correlation and are certainly distinguishable from voids. A visual inspection of tendrils, shown as the green points in the second and fourth panels of Figure \ref{fig:strip}, confirms that they are distinct structures to filaments. Future radio surveys such as ASKAP will help to establish the gas content of tendrils, and if they trace underlying gas flows from voids into filaments.

To understand the impact of tendril galaxies on traditional void classifications, we identify a subset of galaxies in the GLSSC that lie within voids found with a modified voidfinder algorithm applied to the SDSS-DR7 data set, in regions that overlap with the equatorial GAMA fields. We find that only a quarter of these galaxies are determined to be void galaxies in the GLSSC, and that almost 65\% of them are in fact tendril galaxies. We stress the importance of selecting isolated galaxies and voids from the same survey, and note that voids appear to be less empty when observed at fainter magnitudes.

Void galaxies are a necessary tool for understanding the role of secular evolution in the evolution of a galaxy, so it is important to select galaxies that are truly isolated when conducting such studies. Future void galaxy surveys will benefit from a deeper target selection and a high target density in order to identify the most isolated galaxies in the local Universe.

\section*{Acknowledgements}
MA is funded by the University of St Andrews and the International Centre for Radio Astronomy Research. ASGR is supported by funding from a UWA Fellowship. PN acknowledges the support of the Royal Society through the award of a University Research Fellowship and the European Research Council, through receipt of a Starting Grant (DEGAS-259586). MJIB acknowledges the financial support of the Australian Research Council Future Fellowship 100100280. LSK is supported by the Austrian Science Foundation FWF under grant P23946.

GAMA is a joint European-Australasian project based around a spectroscopic campaign using the Anglo-Australian Telescope. The GAMA input catalogue is based on data taken from the Sloan Digital Sky Survey and the UKIRT Infrared Deep Sky Survey. Complementary imaging of the GAMA regions is being obtained by a number of independent survey programs including GALEX MIS, VST KIDS, VISTA VIKING, WISE, Herschel-ATLAS, GMRT and ASKAP providing UV to radio coverage. GAMA is funded by the STFC (UK), the ARC (Australia), the AAO, and the participating institutions. The GAMA website is http://www.gama-survey.org/.

\footnotesize
\bibliographystyle{mn2e}
\setlength{\bibhang}{2.0em}
\setlength{\labelwidth}{0.0em}
\bibliography{tendrils}
\normalsize

\label{lastpage}

\end{document}